\documentclass[conference]{IEEEtran}
\IEEEoverridecommandlockouts

\usepackage{
    amsfonts,
    amsmath,
    array,
    bbding,
    booktabs,
    caption,
    cite,
    color,
    colortbl,
    enumitem,
    float,
    framed,
    graphics,
    graphicx,
    listings,
    makecell,
    multirow,
    subcaption,
    tabularx,
    txfonts,
    upgreek,
    url,
    wrapfig,
    xcolor,
}

\def\BibTeX{{\rm B\kern-.05em{\sc i\kern-.025em b}\kern-.08em
T\kern-.1667em\lower.7ex\hbox{E}\kern-.125emX}}

\newcommand\omni{OmniBioTwin}

\begin{document}

\title{\omni{}: A System-of-Twinned-Systems Framework for Health Digital Twins\\
    \thanks{* Both authors contributed equally to this research.}
}

\author{
    \IEEEauthorblockN{Zhaohui Wang\textsuperscript{*}}
    \IEEEauthorblockA{
        \textit{Indiana University School of Medicine}\\
        Indianapolis, Indiana, USA\\
        zw91@iu.edu
    }
    \and
    \IEEEauthorblockN{Yu Huang\textsuperscript{*}}
    \IEEEauthorblockA{
        \textit{Indiana University School of Medicine}\\
        \textit{Regenstrief Institute}\\
        Indianapolis, Indiana, USA\\
        yh60@iu.edu
    }
    \and
    \IEEEauthorblockN{Jiang Bian}
    \IEEEauthorblockA{
        \textit{Indiana University School of Medicine}\\
        \textit{Regenstrief Institute}\\
        Indianapolis, Indiana, USA\\
        bianji@iu.edu
    }
}

\maketitle

\begin{abstract}
    Health digital twins (HDTs) promise patient-specific modeling and decision support but current approaches remain structurally fragmented: monolithic models that address a single organ or task lack cross-scale fidelity, while system-level twins lack generalizable architectural frameworks. We propose \omni{}, a System-of-Twinned-Systems (SoTS) framework that organizes HDTs as modular computational entities coupled through explicit interaction operators within a multi-layer network architecture. The framework comprises seven coordinated layers---spanning data integration, autonomous twin modeling, cross-scale coupling, temporal synchronization, and human-in-the-loop decision support. We demonstrate \omni{} by instantiating a multiscale twin for glucagon-like peptide-1 (GLP-1) signaling pathways in Alzheimer's disease, illustrating how molecular, cellular, and organ-level twins can be composed and coupled within a unified system.
\end{abstract}


\section{Introduction}
\label{sec:intro}

Health digital twins (HDTs)---dynamic, patient-specific computational representations of health and disease---have emerged as a promising paradigm for precision medicine~\cite{katsoulakis2024digital, willcox2023foundational}.  By integrating diverse data streams longitudinally, HDTs have the potential to support individualized prognosis, simulate therapeutic interventions in silico, and guide clinical decision-making with greater precision.

HDTs have been applied to diverse disease domains, including cardiovascular disease, oncology, immune-mediated disorders, and diabetes~\cite{coorey2022health,wu2022integrating,niarakis2024immune,olawade2025digital}. Yet most implementations remain architecturally monolithic, designed around a specific organ system, biological mechanism, or clinical endpoint. While effective for focused simulation, monolithic frameworks struggle to capture inter-organ communication, cross-scale dependencies, and dynamically evolving interactions among heterogeneous biological processes. In engineering disciplines, System-of-Systems (SoS) digital twin architectures have been proposed to address analogous compositional challenges~\cite{adesanya2025systems}, but these frameworks have not been adapted to the distinct requirements of biomedical modeling, where biological scale heterogeneity, sparse asynchronous observations, and mechanistic interpretability impose additional constraints.

Recent surveys have therefore argued that future HDTs should move beyond isolated organ models toward multi-organ, multiscale, or whole-system representations~\cite{bordukova2024generative,sel2025survey,de2025challenges,ren2025utilization}. Alzheimer's disease (AD) exemplifies the need for such an architecture: as a progressive neurodegenerative disorder shaped by interactions among metabolic dysfunction, vascular impairment, neuroinflammation, and amyloid/tau pathology~\cite{ittner2011amyloid,bloom2014amyloid}, it cannot be adequately represented by single-scale model. Yet no generalizable framework exists for defining biological linkages across scales, coordinating heterogeneous component models, or composing into patient-specific system.

Two developments make a broader HDT framework increasingly feasible. First, multiscale data sources---omics, imaging, wearable streams, and longitudinal electronic health records---are becoming sufficiently rich to support integrated representations of human biology~\cite{peirlinck2021precision,laubenbacher2022building,mosquera2024digital}. Second, emerging SoS-oriented computational paradigms suggest that complex digital representations are better constructed as coordinated assemblies of interacting models rather than as single end-to-end systems. Together, these developments motivate the need for HDT architectures that can support modularity, explicit biological coupling, and flexible integration across scales.

We propose \omni{}, which adapts the SoS paradigm for biomedical HDTs as System-of-Twinned-Systems (SoTS). Each twin represents a specific biological component (e.g., an organ, a cell) or process (e.g., a physiological system), and the full system is organized through explicit mechanisms of communication and coupling across scales. The multi-layer network abstraction~\cite{kivela2014multilayer} provides the structural backbone, with intra-layer edges capturing same-scale interactions and inter-layer edges encoding cross-scale biological coupling. This architecture supports modularity and interpretability while offering a more faithful computational representation of the distributed nature of human disease. In this paper, we describe the \omni{} architecture and demonstrate it through a case study of GLP-1 receptor agonist (GLP-1 RA) pathways in AD, illustrating multiscale twin instantiation and coupling.

\section{Evolution of Architectural Paradigms}

\subsection{Monolithic HDT}

A monolithic HDT represents a biological subsystem within a single level of biological organization, such as a cellular signaling network or an organ-level physiological model. While effective for narrowly scoped tasks, monolithic designs face three structural limitations in systemic diseases such as AD: (1) biological systems span molecular, cellular, tissue, organ, and organism levels, each with distinct state variables and dynamics; (2) these processes evolve on different timescales, from rapid molecular signaling to clinical progression over months or years; and (3) different subsystems may require fundamentally different modeling paradigms---mechanistic, stochastic, or data-driven. Integrating them within a single monolithic framework reduces interpretability and limits independent validation and extension of individual components.

\begin{figure}[htbp]
    \centering
    \includegraphics[width=1\linewidth]{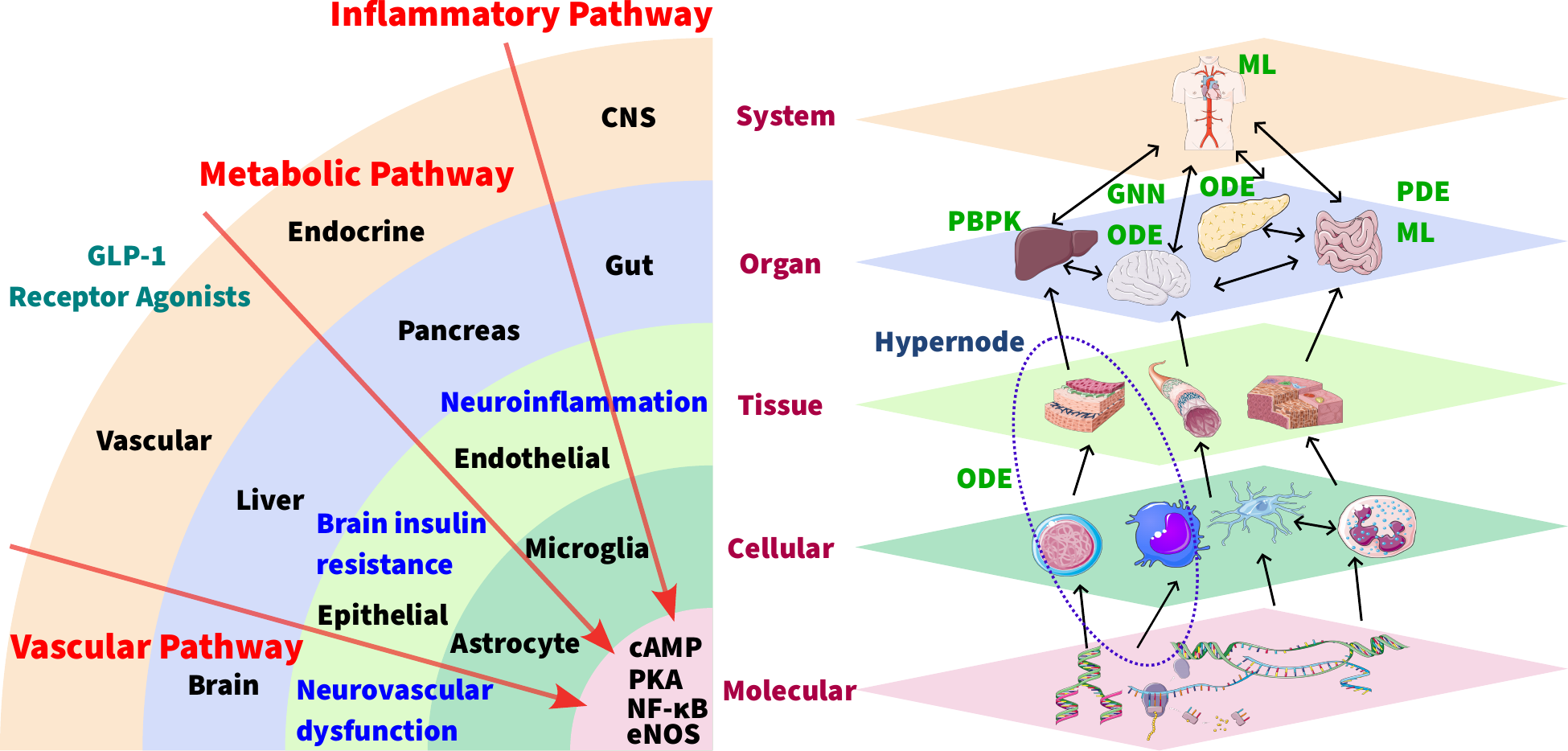}
    \caption{Multi-Layer Network of GLP-1 Signaling.}
    \label{fig:mln}
\end{figure}

\subsection{The System-of-Twinned-Systems Architecture}

To address these limitations, we represent multiscale disease dynamics using a multi-layer network abstraction~\cite{kivela2014multilayer}, as illustrated in Fig~\ref{fig:mln}. Nodes represent biological entities or processes, edges represent interactions, and each layer corresponds to a biological scale (molecular, cellular, tissue, organ, or system). Intra-layer edges capture interactions at similar scales, whereas inter-layer edges encode vertical cross-scale coupling. For example, neuronal metabolic dysfunction may alter vascular function, while vascular impairment may promote inflammatory activation through blood--brain barrier disruption. Because biological systems often involve many-to-many interactions, pairwise edges alone may be insufficient. Inflammatory signaling, for instance, may involve microglia, astrocytes, endothelial cells, and infiltrating immune cells. To capture such collective behavior, we introduce \textit{hypernodes}, which group multiple twins into composite interaction units.

Building on this abstraction, the SoTS paradigm treats each digital twin as an autonomous computational subsystem coupled to others through explicit interaction mechanisms~\cite{adesanya2025systems}. In our biomedical SoTS architecture, each twin maintains its own states, observations, dynamics, and uncertainty representation, and may adopt the modeling strategy most appropriate to the subsystem of interest. For example, a metabolic twin may use a mechanistic biochemical model, whereas an inflammation twin may use a probabilistic state model. Interactions among twins are mediated by explicit coupling operators that propagate data, latent states, causal signals, or uncertainty estimates while preserving interpretability. This modular design also improves scalability, since new twins can be incorporated without redesigning the entire system. The following section details the \omni{} architecture that realizes this paradigm.

\section{Architecture of \omni{}}

\begin{figure*}[t]
    \centering
    \includegraphics[width=1\linewidth]{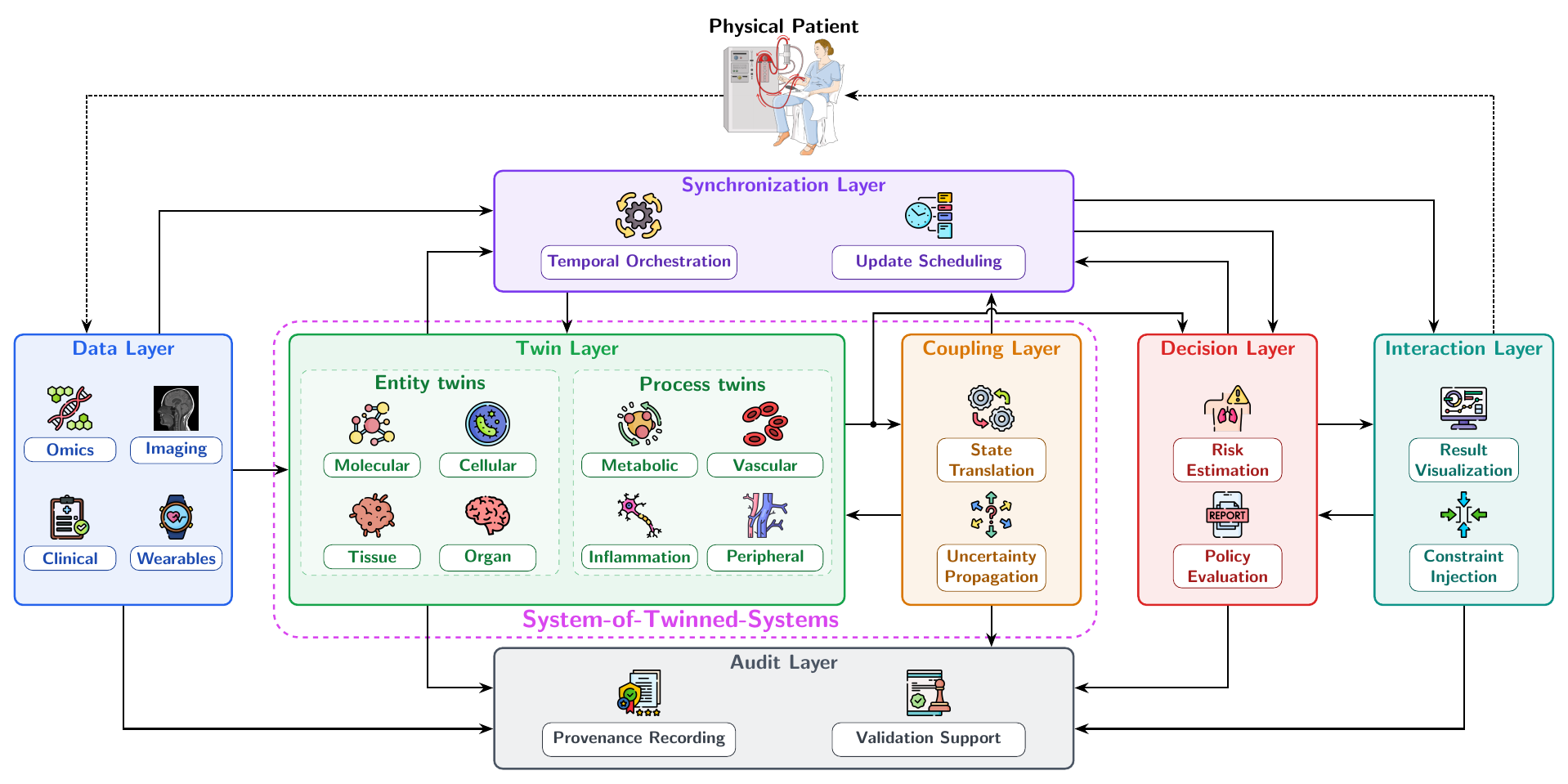}
    \caption{The Architecture of \omni{}.}
    \label{fig:omni_arch}
\end{figure*}

The architecture of \omni{} in Fig~\ref{fig:omni_arch} addresses a central challenge in HDT construction: \textit{how to integrate heterogeneous biomedical data, multi-paradigm models, and dynamic decision processes into a coherent computational framework while preserving biological interpretability and modular extensibility}. At deployment, \omni{} is instantiated as a patient-specific closed-loop digital twin that continuously receives observations from the physical patient and returns interpretable predictions, monitoring summaries, and approved intervention signals to the clinical environment. The architecture is connected through structured data integration, cross-twin coupling, multi-rate synchronization, and human-in-the-loop decision support, enabling information to propagate across biological scales and functional subsystems while supporting ongoing monitoring and adaptive intervention. These functions are realized through seven coordinated layers---\textit{Data}, \textit{Twin}, \textit{Coupling}, \textit{Synchronization}, \textit{Decision}, \textit{Interaction}, and \textit{Audit}---described below.

\subsection{Data Layer}
The \textit{Data Layer} provides the observational foundation of \omni{}, organizing heterogeneous measurements from the physical patient---omics, imaging, clinical records, and continuous monitoring streams---while preserving biological scale, temporal structure, and provenance. It also records realized intervention events such as treatment administration and dosage timing. Each observation is accompanied by metadata and is exposed in a \textit{twin-specific}, rather than globally homogenized form. Each twin $T_i$ defines its own admissible input schema, the \textit{Data Layer} produces an input bundle:
$$
Y_i^t = \left( X_i^t, \tau_i^t, \Sigma_i^t, \Pi_i^t, A_i^t \right)
$$
where $X_i^t$ denotes processed observations, $\tau_i^t$ temporal metadata, $\Sigma_i^t$ measurement uncertainty, $\Pi_i^t$ provenance records, and $A_i^t$ realized intervention records relevant to twin $i$. These outputs are propagated to downstream layers: twin-specific inputs to the \textit{Twin Layer}, temporal metadata and action-timing information to the \textit{Synchronization Layer}, and provenance records to the \textit{Audit Layer}.

\subsection{SoTS in \omni{}}
\subsubsection{Twin Layer}
\label{sec:twin_layer}

The \textit{Twin Layer} contains the core computational units of \omni{} and represents both biological entities and processes across scales. Each twin is modeled as an autonomous computational unit with its own state representation, update mechanism, and uncertainty characterization. Formally, each twin $T_i$ at time $t$ is defined as:
$$
T_i^t=\left(S_i^t,f_i,U_i^t\right)
$$
where $S_i^t$ is the internal state with $S_i^t \in \mathcal{S}_i$, $\mathcal{S}_i$ is the state space, $f_i$ is the transition function, and
$U_i^t$ is the unified uncertainty representation associated with twin $i$. Here,
$$
U_i^t=\left(\Sigma_i^t,\Omega_i^t\right)
$$
with $\Sigma_i^t$ denoting incoming uncertainty measurements and $\Omega_i^t$ denoting uncertainty associated with the internal twin state. The functional form of $f_i$ may be mechanistic, statistical, machine learning-based, or hybrid, depending on the subsystem.

Each twin receives four categories of input: twin-specific processed observations and measurement uncertainty from the \textit{Data Layer}, coupling messages from the \textit{Coupling Layer}, scheduling variables from the \textit{Synchronization Layer}, and intervention information derived from the realized intervention records in the \textit{Data Layer}. Specifically, let
$$
I_i^t = H_i\left(A_i^t\right)
$$
denote the twin-specific intervention representation obtained from the intervention record $A_i^t$ under a mapping $H_i$. The local update proposed by twin $i$ is then:
$$
\left(S_i^{t+1},U_i^{t+1}\right)
=
f_i\left(S_i^t,U_i^t,X_i^t,\Gamma_i^t,I_i^t\right)
$$
where $\Gamma_i^t$ denotes the total cross-twin influence acting on twin $i$ (Section~\ref{sec:coup_layer}). Updated state--uncertainty pairs are then passed to the \textit{Coupling Layer} and the \textit{Decision Layer}, while update signals and logs are propagated to the \textit{Synchronization Layer} and the \textit{Audit Layer}.

\subsubsection{Coupling Layer}
\label{sec:coup_layer}

The \textit{Coupling Layer} formalizes how one twin influences another and serves as the primary mechanism for system-level integration in \omni{}. It preserves twin autonomy while enabling structured propagation of state and uncertainty across biological scales and functional subsystems. Its inputs consist of state--uncertainty pairs $\left(S_i^t,U_i^t\right)$ from the \textit{Twin Layer}. For each ordered pair of interacting twins $\left(j,i\right)$, the layer applies a coupling operator:
$$
C_{j\rightarrow i}\left(S_j^t,U_j^t\right)
$$
which transforms the source twin state into a representation admissible for the target twin. These operators perform cross-scale translation, semantic alignment, unit conversion, and uncertainty propagation. The resulting aggregate coupling message is:
$$
\Gamma_i^t=\sum_{j\neq i} C_{j\rightarrow i}\left(S_j^t,U_j^t\right)
$$
These messages are transmitted back to the \textit{Twin Layer}, while dependency signals are provided to the \textit{Synchronization Layer}. In AD-related applications, such coupling is inherently bidirectional: metabolic dysfunction may induce vascular stress, vascular impairment may amplify inflammatory activation, and inflammatory activity may aggravate both metabolic and vascular dysregulation.

\subsection{Synchronization Layer}

The \textit{Synchronization Layer} provides the temporal orchestration mechanism of \omni{}, determining when and in what order local state updates and coupling interactions are executed. This role is essential in biomedical systems, where observations are sparse, asynchronous, and collected at heterogeneous temporal resolutions. Its inputs include temporal metadata from the \textit{Data Layer}, local update requests from the \textit{Twin Layer}, and dependency signals from the \textit{Coupling Layer}. Based on these inputs, the layer determines whether twin $i$ should be updated at time $t$, via a scheduling variable $\delta_i^t \in \{0,1\}$. The resulting update schedules and synchronized state trajectories are propagated to downstream layers. In AD settings, this mechanism coordinates processes evolving on unequal timescales, from rapid molecular signaling to slower imaging and cognitive progression.

\subsection{Decision Layer}

The \textit{Decision Layer} serves as the system-level inference and policy engine of \omni{}. It transforms synchronized state and uncertainty information into clinically or scientifically actionable outputs, including individualized risk estimates, treatment-response assessments, and candidate intervention or monitoring recommendations. Its inputs consist of current state--uncertainty pairs from the \textit{Twin Layer}, synchronized trajectories from the \textit{Synchronization Layer}, together with optional human-provided constraints from the \textit{Interaction Layer}. At the system level, the decision process operates on the synchronized global state:
$$
\left(\widetilde{\mathbf{S}}^t,\widetilde{\mathbf{U}}^t\right)
=
\left(\{S_i^t\}_{i=1}^N,\{U_i^t\}_{i=1}^N\right)
$$
and produces a pair of outputs:
$$
\left(O^t,\hat{A}^t\right)
=
D\left(
    \widetilde{\mathbf{S}}^{0:t},\widetilde{\mathbf{U}}^{0:t},c^t
\right)
$$
Here, $\widetilde{\mathbf{S}}^{0:t}$ and $\widetilde{\mathbf{U}}^{0:t}$ denote synchronized histories, and $c^t$ denotes external constraints from the \textit{Interaction Layer}. The output $O^t$ represents predictive summaries such as risk scores, projected outcomes, or uncertainty-aware trajectory assessments, whereas $\hat{A}^t$ denotes a candidate intervention, monitoring recommendation, or other proposed action. These outputs are transmitted to the \textit{Interaction Layer} and the \textit{Audit Layer}.

\subsection{Interaction Layer}

The \textit{Interaction Layer} is the human-facing interface of \omni{}, responsible for rendering model behavior interpretable and operationally usable for clinicians, researchers, and domain experts. It translates synchronized trajectories and decision summaries into explanations, plots, and visualizations. It also serves as the interface through which users may specify constraints for the \textit{Decision Layer} and review or approve candidate actions proposed by it. Approved actions are then communicated from the \textit{Interaction Layer} to the external physical patient, while records of user interactions are transmitted to the \textit{Audit Layer}. When such actions are later realized in practice, their execution and adherence information re-enter \omni{} through the \textit{Data Layer} as part of the observed input stream.

\subsection{Audit Layer}

The \textit{Audit Layer} provides the traceability, validation, and governance infrastructure of \omni{}. It aggregates provenance records from the \textit{Data Layer}, state and uncertainty trajectories from the \textit{Twin Layer}, coupling traces from the \textit{Coupling Layer}, synchronization logs, decision outputs, and user interaction records. These are consolidated into structured audit records, traceability reports, and validation documentation. In this way, the \textit{Audit Layer} ensures that \omni{} is not only predictive and multiscale, but also reproducible, explainable, and suitable for biomedical validation and eventual regulatory assessment.

\section{GLP-1 Pathways in AD}

As introduced in Section~\ref{sec:intro}, AD reflects interactions among multiple biological processes---neuronal metabolism, neurovascular dysfunction, neuroinflammation, and amyloid/tau pathology~\cite{ittner2011amyloid,bloom2014amyloid}---with marked inter-patient heterogeneity that motivates mechanism-specific and patient-specific modeling (Fig~\ref{fig:mln}). Within this context, GLP-1 RAs provide an informative pharmacological probe for multiscale AD modeling, since their effects extend beyond glycemic control and engage several interacting AD-relevant processes.

\subsection{Metabolic Pathway in AD}
\label{sec:meta}

Within the metabolic axis, GLP-1 RAs modulate neuronal insulin signaling and bioenergetic resilience. AD brains commonly exhibit reduced cerebral glucose metabolism, often reflected by FDG-PET hypometabolism, together with molecular evidence consistent with impaired brain insulin signaling, including altered IR--IRS-1--PI3K signaling and increased inhibitory serine phosphorylation of IRS-1~\cite{talbot2012demonstrated}. A$\upbeta$ oligomers have been shown to aggravate this defect, in part through JNK- and TNF-$\upalpha$-related signaling, thereby weakening physiological insulin signaling~\cite{bomfim2012anti}. In this setting, GLP-1 receptor activation has been reported to engage cAMP--PKA--CREB signaling~\cite{holscher2014central}, support neuronal survival and synaptic plasticity through downstream mediators such as BDNF~\cite{ohtake2014exendin,mcclean2011diabetes}, and activate PI3K--Akt signaling while inhibiting GSK-3$\upbeta$, a key contributor to tau hyperphosphorylation and A$\upbeta$-associated toxicity~\cite{wang2018glp}. 

\subsection{Instantiation in the \omni{} Framework}

For therapeutic modeling in living patients with AD, the \omni{} framework builds a set of coupled twins using clinically accessible, non-invasive data and infers latent disease states. This design is motivated by the fact that AD is not a single-scale disorder.

The \textit{Peripheral Twin} is constructed from treatment records and routine peripheral measurements. Its inputs include GLP-1 RA intervention records together with plasma biomarkers such as glucose, insulin, and HbA1c, as well as body weight or BMI~\cite{fu2020brain,lara2025multiscale}. Using these data, the \textit{Peripheral Twin} characterizes systemic drug exposure and peripheral metabolic response and outputs latent peripheral states such as effective peripheral GLP-1 RA exposure, glycemic response, systemic insulin sensitivity, metabolic stress, and a brain-directed GLP-1 signaling drive~\cite{fu2020brain}. In this way, the \textit{Peripheral Twin} serves as the entry point of the intervention into the system.

The \textit{Molecular Twin} takes this brain-directed GLP-1 signaling drive, together with glycemic response and systemic insulin sensitivity from the \textit{Peripheral Twin}, and combines them with non-invasive molecular observations, including CSF biomarkers such as A$\upbeta$42/A$\upbeta$40 and p-tau181/p-tau217, plasma biomarkers such as p-tau217, A$\upbeta$42/A$\upbeta$40, NfL, and GFAP, and molecular imaging measurements including amyloid PET, tau PET, and FDG-PET~\cite{lara2025multiscale}. Based on these inputs, the \textit{Molecular Twin} reconstructs latent AD-relevant molecular pathology, including A$\upbeta$ burden~\cite{roh2026evolving}, tau phosphorylation pressure~\cite{ashton2025alzheimer}, neuroinflammatory activation~\cite{leipp2024glial}, and brain insulin resistance or metabolic dysfunction~\cite{frisoni2025new}. These outputs convert treatment-related and biomarker-derived information into biologically meaningful molecular pressures that can act on the cellular layer.

The \textit{Cellular Twin} is built on the outputs of the \textit{Molecular Twin}. Its inputs are molecularly translated pressures, including A$\upbeta$ toxicity pressure, tau toxicity pressure, neuroinflammatory activation, and metabolic support level. These quantities are used to model how major brain cell populations, including neurons, astrocytes, and microglia, change state and interact under AD-related stress. The \textit{Cellular Twin} then outputs cell-level states such as neuronal functional integrity~\cite{rather2024influence}, synaptic density~\cite{mcgeachan2025amyloid}, and glial activation state, including both microglial and astrocytic reactivity~\cite{deng2024microglia}, as well as overall cellular vulnerability. In this way, the \textit{Cellular Twin} translates molecular pathology into functional cell-level consequences.

The \textit{Organ Twin} receives these cell-derived regional summaries from the \textit{Cellular Twin}, specifically synaptic density, neuronal integrity, and inflammation load, and combines them with organ-level observations from structural and functional neuroimaging and cognitive testing. Its inputs therefore include MRI measures of brain atrophy, DTI measures of structural connectivity~\cite{lombardi2020association}, fMRI measures of functional connectivity, FDG-PET measures of brain metabolism, and cognitive scores such as MMSE and ADAS-Cog. Using these inputs, the \textit{Organ Twin} models how cell-level dysfunction accumulates across brain regions and networks to generate macroscopic disease progression. Its outputs include brain network dysfunction, functional connectivity disruption, cognitive decline trajectory~\cite{badhwar2020multiomics}, and regional atrophy trajectory~\cite{lara2025multiscale}. Thus, the \textit{Organ Twin} converts regional cellular injury into clinically relevant whole-brain phenotypes that can be compared directly with imaging and cognitive readouts in living patients.

Taken together, the four twins define a multiscale progression from treatment and systemic physiology to latent molecular pathology, then to cellular dysfunction, and finally to organ-level degeneration and cognitive decline. On this basis, the \textit{Decision Layer} integrates the outputs of all four twins to generate patient-specific summaries, including predicted disease progression, estimated response to GLP-1 RA treatment, and candidate recommendations for therapy adjustment or follow-up monitoring. The \textit{Interaction Layer} then presents these results in an interpretable form, such as cross-scale trajectory summaries and counterfactual comparisons, allowing clinicians to review, constrain, and approve suggested actions. Approved interventions are subsequently applied to the patient and re-enter the system through updated clinical observations, thereby closing the loop between multiscale inference and clinical management.

\subsection{Potential Data Sources}

The construction of a multiscale digital twin in living patients requires the integration of heterogeneous real-world data sources. At the clinical level, electronic health records (EHRs) provide longitudinal information on demographics, comorbidities, medication history, laboratory measurements, and clinical outcomes. These data are particularly useful for constructing the \textit{Peripheral Twin} and informing the \textit{Organ Twin}, as they capture treatment exposure, metabolic status, and disease progression over time.

At the molecular and cellular levels, high-throughput omics datasets provide key information on disease mechanisms. Public resources such as ssREAD~\cite{wang2024single} offer cell-type-resolved and spatially localized gene expression profiles across brain regions and disease stages. These data can be used to calibrate the \textit{Molecular Twin} and \textit{Cellular Twin}, particularly for pathway activity, cell-type-specific responses, and spatial heterogeneity.

At the organ level, neuroimaging datasets such as ADNI~\cite{jack2008alzheimer} provide MRI, DTI, fMRI, and PET measurements that are essential for constructing and validating the \textit{Organ Twin}, as they reflect brain-wide atrophy, connectivity changes, and metabolic dysfunction.

\section{Challenges and Future Work}
Although \omni{} provides a structured architecture for multiscale HDTs, important challenges remain before such systems can become scientifically robust and clinically useful. These include: (1) multimodal alignment and uncertainty-aware harmonization across omics, imaging, clinical, and wearable data; (2) the design of sparse, biologically grounded, and stable coupling operators across interacting twins; (3) principled propagation and calibration of uncertainty at both twin and system levels; (4) empirical validation demonstrating that modular SoTS architectures improve predictive fidelity or clinical utility over monolithic alternatives; and (5) governance requirements including privacy-preserving integration, consent-aware oversight, fairness assessment, and accountability in clinical decision-support settings.

\section{Conclusion}

In this work, we introduced \omni{}, a \textbf{System-of-Twinned-Systems} framework in which digital twins are treated as modular computational entities coupled through an explicit multi-layer architecture. By adapting the SoS digital twin paradigm---previously confined to engineering domains---to biomedical modeling, \omni{} provides a scalable and interpretable foundation for integrating heterogeneous data, modeling paradigms, and biological processes. The GLP-1/AD case study illustrates how twins can be instantiated and coupled within this architecture, grounding the framework in a concrete multiscale disease context.

\section*{Acknowledgment}
This study was supported in part by National Institutes of Health grants R01AG089445, U01AG088076, R01AG083039, RF1AG084178, R01AG084236, RF1AG077820, R01AG080991, R01AG080624, and R01AG076234.

\bibliographystyle{ieeetr}
\bibliography{reference}

@article{peirlinck2021precision,
  title         = {Precision medicine in human heart modeling: Perspectives, challenges, and opportunities},
  author        = {Peirlinck, M and Costabal, F Sahli and Yao, J and Guccione, JM and Tripathy, S and Wang, Y and Ozturk, D and Segars, P and Morrison, TM and Levine, S and others},
  journal       = {Biomechanics and modeling in mechanobiology},
  volume        = {20},
  number        = {3},
  pages         = {803--831},
  year          = {2021},
  publisher     = {Springer}
}

@article{laubenbacher2022building,
  title         = {Building digital twins of the human immune system: toward a roadmap},
  author        = {Laubenbacher, Reinhard and Niarakis, Anna and Helikar, Tom{\'a}{\v{s}} and An, Gary and Shapiro, Bruce and Malik-Sheriff, Rahuman S and Sego, TJ and Knapp, Adam and Macklin, Paul and Glazier, James A},
  journal       = {NPJ digital medicine},
  volume        = {5},
  number        = {1},
  pages         = {64},
  year          = {2022},
  publisher     = {Nature Publishing Group UK London}
}

@article{wu2022integrating,
  title         = {Integrating mechanism-based modeling with biomedical imaging to build practical digital twins for clinical oncology},
  author        = {Wu, Chengyue and Lorenzo, Guillermo and Hormuth, David A and Lima, Ernesto ABF and Slavkova, Kalina P and DiCarlo, Julie C and Virostko, John and Phillips, Caleb M and Patt, Debra and Chung, Caroline and others},
  journal       = {Biophysics reviews},
  volume        = {3},
  number        = {2},
  year          = {2022},
  publisher     = {AIP Publishing}
}

@article{katsoulakis2024digital,
  title         = {Digital twins for health: a scoping review},
  author        = {Katsoulakis, Evangelia and Wang, Qi and Wu, Huanmei and Shahriyari, Leili and Fletcher, Richard and Liu, Jinwei and Achenie, Luke and Liu, Hongfang and Jackson, Pamela and Xiao, Ying and others},
  journal       = {NPJ digital medicine},
  volume        = {7},
  number        = {1},
  pages         = {77},
  year          = {2024},
  publisher     = {Nature Publishing Group UK London}
}

@article{coorey2022health,
  title         = {The health digital twin to tackle cardiovascular disease--a review of an emerging interdisciplinary field},
  author        = {Coorey, Genevieve and Figtree, Gemma A and Fletcher, David F and Snelson, Victoria J and Vernon, Stephen Thomas and Winlaw, David and Grieve, Stuart M and McEwan, Alistair and Yang, Jean Yee Hwa and Qian, Pierre and others},
  journal       = {NPJ digital medicine},
  volume        = {5},
  number        = {1},
  pages         = {126},
  year          = {2022},
  publisher     = {Nature Publishing Group UK London}
}

@article{bordukova2024generative,
  title         = {Generative artificial intelligence empowers digital twins in drug discovery and clinical trials},
  author        = {Bordukova, Maria and Makarov, Nikita and Rodriguez-Esteban, Raul and Schmich, Fabian and Menden, Michael P},
  journal       = {Expert opinion on drug discovery},
  volume        = {19},
  number        = {1},
  pages         = {33--42},
  year          = {2024},
  publisher     = {Taylor \& Francis}
}

@book{willcox2023foundational,
  title         = {Foundational research gaps and future directions for digital twins},
  author        = {Willcox, Karen and Bingham, D and Chung, C and Chung, J and Cruz-Neira, C and Grant, C and Kinter, J and Leung, R and Moin, P and Ohno-Machado, L and others},
  year          = {2023},
  publisher     = {National Academies Press Washington, DC, USA}
}

@article{mosquera2024digital,
  title         = {Digital twins and artificial intelligence in metabolic disease research},
  author        = {Mosquera-Lopez, Clara and Jacobs, Peter G},
  journal       = {Trends in Endocrinology \& Metabolism},
  volume        = {35},
  number        = {6},
  pages         = {549--557},
  year          = {2024},
  publisher     = {Elsevier}
}

@article{niarakis2024immune,
  title         = {Immune digital twins for complex human pathologies: applications, limitations, and challenges},
  author        = {Niarakis, Anna and Laubenbacher, Reinhard and An, Gary and Ilan, Yaron and Fisher, Jasmin and Flobak, {\AA}smund and Reiche, Kristin and Rodr{\'\i}guez Mart{\'\i}nez, Mar{\'\i}a and Geris, Liesbet and Ladeira, Luiz and others},
  journal       = {NPJ systems biology and applications},
  volume        = {10},
  number        = {1},
  pages         = {141},
  year          = {2024},
  publisher     = {Nature Publishing Group UK London}
}

@article{sel2025survey,
  title         = {Survey and perspective on verification, validation, and uncertainty quantification of digital twins for precision medicine},
  author        = {Sel, Kaan and Hawkins-Daarud, Andrea and Chaudhuri, Anirban and Osman, Deen and Bahai, Ahmad and Paydarfar, David and Willcox, Karen and Chung, Caroline and Jafari, Roozbeh},
  journal       = {npj Digital Medicine},
  volume        = {8},
  number        = {1},
  pages         = {40},
  year          = {2025},
  publisher     = {Nature Publishing Group UK London}
}

@article{olawade2025digital,
  title         = {Digital twin paradigm in diabetes prediction and management},
  author        = {Olawade, David B and Owhonda, Rita Chikeru and Alabi, John Oluwatosin and Egbon, Eghosasere and Daniel, Raphael Igbarumah Ayo and Bello, Oluwakemi Jumoke},
  journal       = {Diabetes research and clinical practice},
  pages         = {113075},
  year          = {2025},
  publisher     = {Elsevier}
}

@article{de2025challenges,
  title         = {Challenges and opportunities for digital twins in precision medicine from a complex systems perspective},
  author        = {De Domenico, Manlio and Allegri, Luca and Caldarelli, Guido and d'Andrea, Valeria and Di Camillo, Barbara and Rocha, Luis M and Rozum, Jordan and Sbarbati, Riccardo and Zambelli, Francesco},
  journal       = {npj Digital Medicine},
  volume        = {8},
  number        = {1},
  pages         = {37},
  year          = {2025},
  publisher     = {Nature Publishing Group UK London}
}

@article{ren2025utilization,
  title         = {Utilization of precision medicine digital twins for drug discovery in Alzheimer's disease},
  author        = {Ren, Yunxiao and Pieper, Andrew A and Cheng, Feixiong},
  journal       = {Neurotherapeutics},
  volume        = {22},
  number        = {3},
  pages         = {e00553},
  year          = {2025},
  publisher     = {Elsevier}
}

@article{adesanya2025systems,
  title         = {Systems of twinned systems: A systematic literature review},
  author        = {Adesanya, Feyi and Silva, Kanan Castro and Neto, Valdemar V Graciano and David, Istvan},
  journal       = {arXiv preprint arXiv:2505.19916},
  year          = {2025}
}

@article{talbot2012demonstrated,
  title         = {Demonstrated brain insulin resistance in Alzheimer's disease patients is associated with IGF-1 resistance, IRS-1 dysregulation, and cognitive decline},
  author        = {Talbot, Konrad and Wang, Hoau-Yan and Kazi, Hala and Han, Li-Ying and Bakshi, Kalindi P and Stucky, Andres and Fuino, Robert L and Kawaguchi, Krista R and Samoyedny, Andrew J and Wilson, Robert S and others},
  journal       = {The Journal of clinical investigation},
  volume        = {122},
  number        = {4},
  pages         = {1316--1338},
  year          = {2012},
  publisher     = {American Society for Clinical Investigation}
}

@article{bomfim2012anti,
  title         = {An anti-diabetes agent protects the mouse brain from defective insulin signaling caused by Alzheimer's disease--associated A$\beta$ oligomers},
  author        = {Bomfim, Theresa R and Forny-Germano, Leticia and Sathler, Luciana B and Brito-Moreira, Jordano and Houzel, Jean-Christophe and Decker, Helena and Silverman, Michael A and Kazi, Hala and Melo, Helen M and McClean, Paula L and others},
  journal       = {The Journal of clinical investigation},
  volume        = {122},
  number        = {4},
  pages         = {1339--1353},
  year          = {2012},
  publisher     = {American Society for Clinical Investigation}
}

@article{mcclean2011diabetes,
  title         = {The diabetes drug liraglutide prevents degenerative processes in a mouse model of Alzheimer's disease},
  author        = {McClean, Paula L and Parthsarathy, Vadivel and Faivre, Emilie and H{\"o}lscher, Christian},
  journal       = {Journal of Neuroscience},
  volume        = {31},
  number        = {17},
  pages         = {6587--6594},
  year          = {2011},
  publisher     = {Society for Neuroscience}
}

@article{holscher2014central,
  title         = {Central effects of GLP-1: new opportunities for treatments of neurodegenerative diseases},
  author        = {H{\"o}lscher, Christian},
  journal       = {Journal of Endocrinology},
  volume        = {221},
  number        = {1},
  pages         = {T31--T41},
  year          = {2014},
  publisher     = {Bioscientifica Ltd}
}

@article{ohtake2014exendin,
  title         = {Exendin-4 promotes the membrane trafficking of the AMPA receptor GluR1 subunit and ADAM10 in the mouse neocortex},
  author        = {Ohtake, Nobuaki and Saito, Mieko and Eto, Masaaki and Seki, Kenjiro},
  journal       = {Regulatory peptides},
  volume        = {190},
  pages         = {1--11},
  year          = {2014},
  publisher     = {Elsevier}
}

@article{wang2018glp,
  title         = {GLP-1 receptor agonists downregulate aberrant GnT-III expression in Alzheimer's disease models through the Akt/GSK-3$\beta$/$\beta$-catenin signaling},
  author        = {Wang, Ying and Chen, Song and Xu, Zheng and Chen, Suting and Yao, Wenbing and Gao, Xiangdong},
  journal       = {Neuropharmacology},
  volume        = {131},
  pages         = {190--199},
  year          = {2018},
  publisher     = {Elsevier}
}

@article{kivela2014multilayer,
  title         = {Multilayer networks},
  author        = {Kivel{\"a}, Mikko and Arenas, Alex and Barthelemy, Marc and Gleeson, James P and Moreno, Yamir and Porter, Mason A},
  journal       = {Journal of complex networks},
  volume        = {2},
  number        = {3},
  pages         = {203--271},
  year          = {2014},
  publisher     = {Oxford University Press}
}

@article{ittner2011amyloid,
  title         = {Amyloid-$\beta$ and tau--a toxic pas de deux in Alzheimer's disease},
  author        = {Ittner, Lars M and G{\"o}tz, J{\"u}rgen},
  journal       = {Nature Reviews Neuroscience},
  volume        = {12},
  number        = {2},
  pages         = {67--72},
  year          = {2011},
  publisher     = {Nature Publishing Group UK London}
}

@article{bloom2014amyloid,
  title         = {Amyloid-$\beta$ and tau: the trigger and bullet in Alzheimer disease pathogenesis},
  author        = {Bloom, George S},
  journal       = {JAMA neurology},
  volume        = {71},
  number        = {4},
  pages         = {505--508},
  year          = {2014}
}

@article{fu2020brain,
  title         = {Brain endothelial cells regulate glucagon-like peptide 1 entry into the brain via a receptor-mediated process},
  author        = {Fu, Zhuo and Gong, Liying and Liu, Jia and Wu, Jing and Barrett, Eugene J and Aylor, Kevin W and Liu, Zhenqi},
  journal       = {Frontiers in physiology},
  volume        = {11},
  pages         = {555},
  year          = {2020},
  publisher     = {Frontiers Media SA}
}

@article{badhwar2020multiomics,
  title         = {A multiomics approach to heterogeneity in Alzheimer's disease: focused review and roadmap},
  author        = {Badhwar, AmanPreet and McFall, G Peggy and Sapkota, Shraddha and Black, Sandra E and Chertkow, Howard and Duchesne, Simon and Masellis, Mario and Li, Liang and Dixon, Roger A and Bellec, Pierre},
  journal       = {Brain},
  volume        = {143},
  number        = {5},
  pages         = {1315--1331},
  year          = {2020},
  publisher     = {Oxford University Press}
}

@article{lara2025multiscale,
  title         = {Multiscale networks in Alzheimer's disease identify brain hypometabolism as central across biological scales},
  author        = {Lara-Simon, Elena and Gispert, Juan Domingo and Garcia-Ojalvo, Jordi and Villoslada, Pablo and Alzheimer's Disease Neuroimaging Initiative},
  journal       = {PLOS Computational Biology},
  volume        = {21},
  number        = {10},
  pages         = {e1013583},
  year          = {2025},
  publisher     = {Public Library of Science San Francisco, CA USA}
}

@article{roh2026evolving,
  title         = {Evolving Alzheimer's Disease Clinical Practice: Updated Diagnostic Criteria, Fluid Biomarkers, and Special Considerations for Anti-Amyloid Therapies},
  author        = {Roh, Hyun Woong and Chang, Yoon Young and Kim, Keun You and Jeon, So Yeon and Wang, Sheng-Min and Kim, Eosu and Bae, Jae-Nam and Ryu, Seung-Ho},
  journal       = {Psychiatry Investigation},
  volume        = {23},
  number        = {2},
  pages         = {183},
  year          = {2026}
}

@article{ashton2025alzheimer,
  title         = {The Alzheimer's association global biomarker standardization consortium (GBSC) plasma phospho-tau round robin study},
  author        = {Ashton, Nicholas J and Keshavan, Ashvini and Brum, Wagner S and Andreasson, Ulf and Arslan, Burak and Droescher, Mathias and Barghorn, Stefan and Vanbrabant, Jeroen and Lambrechts, Charlotte and Van Loo, Maxime and others},
  journal       = {Alzheimer's \& Dementia},
  volume        = {21},
  number        = {2},
  pages         = {e14508},
  year          = {2025},
  publisher     = {Wiley Online Library}
}

@article{leipp2024glial,
  title         = {Glial fibrillary acidic protein in Alzheimer's disease: a narrative review},
  author        = {Leipp, Florine and Vialaret, J{\'e}r{\^o}me and Mohaupt, Pablo and Coppens, Salom{\'e} and Jaffuel, Aurore and Niehoff, Ann-Christin and Lehmann, Sylvain and Hirtz, Christophe},
  journal       = {Brain Communications},
  volume        = {6},
  number        = {6},
  pages         = {fcae396},
  year          = {2024},
  publisher     = {Oxford University Press UK}
}

@article{frisoni2025new,
  title         = {New landscape of the diagnosis of Alzheimer's disease},
  author        = {Frisoni, Giovanni B and Hansson, Oskar and Nichols, Emma and Garibotto, Valentina and Schindler, Suzanne E and van der Flier, Wiesje M and Jessen, Frank and Villain, Nicolas and Arenaza-Urquijo, Eider M and Crivelli, Lucia and others},
  journal       = {The Lancet},
  volume        = {406},
  number        = {10510},
  pages         = {1389--1407},
  year          = {2025},
  publisher     = {Elsevier}
}

@article{deng2024microglia,
  title         = {Microglia and astrocytes in Alzheimer's disease: significance and summary of recent advances},
  author        = {Deng, Qianting and Wu, Chongyun and Parker, Emily and Liu, Timon Cheng-Yi and Duan, Rui and Yang, Luodan},
  journal       = {Aging and disease},
  volume        = {15},
  number        = {4},
  pages         = {1537},
  year          = {2024}
}

@article{rather2024influence,
  title         = {Influence of tau on neurotoxicity and cerebral vasculature impairment associated with Alzheimer's disease},
  author        = {Rather, Mashoque Ahmad and Khan, Andleeb and Jahan, Sadaf and Siddiqui, Arif Jamal and Wang, Lianchun},
  journal       = {Neuroscience},
  volume        = {552},
  pages         = {1--13},
  year          = {2024},
  publisher     = {Elsevier}
}

@article{mcgeachan2025amyloid,
  title         = {Amyloid-Beta Pathology Increases Synaptic Engulfment by Glia in Feline Cognitive Dysfunction Syndrome: A Naturally Occurring Model of Alzheimer's Disease},
  author        = {McGeachan, Robert I and Ewbank, Lucy and Watt, Meg and Sordo, Lorena and Malbon, Alexandra and Salamat, Muhammad Khalid F and Tzioras, Makis and De Frias, Joao Miguel and Tulloch, Jane and Houston, Fiona and others},
  journal       = {European Journal of Neuroscience},
  volume        = {62},
  number        = {3},
  pages         = {e70180},
  year          = {2025},
  publisher     = {Wiley Online Library}
}

@article{lombardi2020association,
  title         = {Association between structural connectivity and generalized cognitive spectrum in Alzheimer's disease},
  author        = {Lombardi, Angela and Amoroso, Nicola and Diacono, Domenico and Monaco, Alfonso and Logroscino, Giancarlo and De Blasi, Roberto and Bellotti, Roberto and Tangaro, Sabina},
  journal       = {Brain sciences},
  volume        = {10},
  number        = {11},
  pages         = {879},
  year          = {2020},
  publisher     = {MDPI}
}

@article{wang2024single,
  title         = {A single-cell and spatial RNA-seq database for Alzheimer's disease (ssREAD)},
  author        = {Wang, Cankun and Acosta, Diana and McNutt, Megan and Bian, Jiang and Ma, Anjun and Fu, Hongjun and Ma, Qin},
  journal       = {Nature Communications},
  volume        = {15},
  number        = {1},
  pages         = {4710},
  year          = {2024},
  publisher     = {Nature Publishing Group UK London}
}

@article{jack2008alzheimer,
  title         = {The Alzheimer's disease neuroimaging initiative (ADNI): MRI methods},
  author        = {Jack Jr, Clifford R and Bernstein, Matt A and Fox, Nick C and Thompson, Paul and Alexander, Gene and Harvey, Danielle and Borowski, Bret and Britson, Paula J and L. Whitwell, Jennifer and Ward, Chadwick and others},
  journal       = {Journal of Magnetic Resonance Imaging: An Official Journal of the International Society for Magnetic Resonance in Medicine},
  volume        = {27},
  number        = {4},
  pages         = {685--691},
  year          = {2008},
  publisher     = {Wiley Online Library}
}

\end{document}